\documentclass[fontsize=9pt,DIV=15, a4paper, twocolumn]{scrartcl}
\usepackage{scrhack}
\usepackage[dvips]{graphicx}
\usepackage{subfigure}
\usepackage[utf8]{inputenc}
\usepackage[T1]{fontenc}
\usepackage{textcomp}
\usepackage{lmodern}
\usepackage{amsmath}
\usepackage{amssymb}
\usepackage[english]{babel}
\usepackage{doi}
\usepackage[style=authoryear,eprint=true,url=false,backend=bibtex,citestyle=authoryear,maxbibnames=8,maxcitenames=1,useprefix=true]{biblatex}

\usepackage{hyperref}

\hypersetup{  
  pdfinfo={  
    Title={Thermal creep assisted dust lifting on Mars: Wind tunnel experiments for the entrainment threshold velocity},  
    Author={Markus Küpper and Gerhard Wurm},
    Subject={}, 
    Keywords={},
    displaydoctitle=true,
  },  
  }

\newcommand{\citep}[1]{(\cite{#1})}
\newcommand{\citet}[1]{\cite{#1}} 
\begin{document}

\title{Thermal creep assisted dust lifting on Mars:\\ Wind tunnel experiments for the entrainment threshold velocity}
\author{ Markus K{\"u}pper, Gerhard Wurm \thanks{Fakult{\"a}t f{\"u}r Physik, Universit{\"a}t Duisburg-Essen, Lotharstr. 1, 47057 Duisburg, Germany \newline Email: markus.kuepper.86@uni-due.de}}
\date{August 1, 2015}
\maketitle
\begin{abstract}
In this work we present laboratory measurements on the reduction of the threshold friction velocity necessary for lifting dust if the dust bed is illuminated. Insolation of a porous soil establishes a temperature gradient. At low ambient pressure this gradient leads to thermal creep gas flow within the soil. This flow leads to a sub-surface overpressure which supports lift imposed by wind. The wind tunnel was run with Mojave Mars Simulant and air at 3, 6 and 9\,mbar, to cover most of the pressure range at martian surface levels. Our first measurements imply that the insolation of the martian surface can reduce the entrainment threshold velocity between 4\,\% and 19\,\% for the conditions sampled with our experiments. An insolation activated soil might therefore provide additional support for aeolian particle transport at low wind speeds.
\end{abstract}

\section{Introduction}

Sand and dust are ubiquitous at the surface of Mars and in its atmosphere. Regional as well as global dust storms are regularly observed \citep{Cantor2007}. In general, the atmospheric dust is thought to play a major role for the martian climate  \citep{fedorova2014,montabone2014,kahre2013}. On the ground, eolian features are wide spread. Recently, \citet{Bridges2012} reported that sand fluxes at the Nili Patera dune field on Mars are comparable to sand fluxes found on earth as the dunes move in almost steady state. This suggests that the underlying physics on Mars and Earth are similiar and that particle lift is related to saltation. However, at the low surface pressure moving particles by wind drag only is far from trivial. In earlier work \citet{greeley1980} showed that most of the time the wind speed on Mars should be too low to move particles.

Dust devils provide an additional view on particle lift due to the pressure drop in their center \citep{balme2006}. The pressure difference between lower atmospheric pressure and higher sub-surface pressure within the soil might support particle lift. Still, \citet{Neakrase2010} reported a significant discrepancy by a factor 3 between the observed mass fluxes and the estimation of dust entrainment at Gusev crater. Seen globally, the lifting related to dust storms might also dominate over lift by dust devils \citep{Whelley2008}. 

To compensate for this difference between observed dust flux and standard lifting capabilities of wind and dust devils, several further ideas were proposed. Certain types of particle inventory like volcanic glass or aggregated particles are more susceptible for pick up at lower wind speeds. The big, light-weight particles can roll over the surfaces which breaks the adhesion before the particles are then  lifted \citep{devet2014,merrison2007}. Charging has also been discussed as a mechanism to explain the high entrainment rates \citep{Kok2006}. If a number of particles are ejected at low wind speed the threshold wind speed will also be lowered due to the reimpacting grains. Steady state saltation might  then occur at lower wind speeds \citep{Kok2010}. For a review on sand and dust entrainment we refer to \citet{Merrison2012}.

It is somewhat counter intuitively, that owing to the low pressure  "new" lifting mechanisms can be strong on Mars which are negligible at the 1\,bar atmosphere on Earth. The mere insolation of a dust bed and related temperature gradients within the porous soil lead to strong lifting forces. This was already shown in \cite{Wurm2008}, though the underlying model changed somewhat (see below). In any case the insolation will "activate" the top soil layer. Thermal creep gas flow within the soil results from the temperature distribution and leads to sub-surface overpressure and therefore, as in the case suggested for dust devils, to an additional lift. This overpressure reduces the erosion threshold wind speed and enhances the entrainment rate. At specific geologic sites additional gas flows through the porous ground occur \citep{Antoine2011,Lopez2012}. This might also add a lifting force.

Here, we report on wind tunnel experiments where we measure for the first time the reduction of the threshold friction velocity in dependence of an insolating light flux. We start by combining the illumination based lift with an aeolian lift model. We then describe the wind tunnel, the particle sample and experiments on particle lift. The data are used to give a measurement of the boundary layer profile and the likelihood for detachment in a light flux / wind velocity parameter set. We apply the lift model to this data set to obtain a threshold friction velocity depending light flux. This model fit is then scaled to martian conditions. It is used to quantify the reduction in friction velocity needed to lift particles on Mars due to insolation.

\subsection{Insolation Activated Layer}

In recent years a number of papers showed that at low pressure 
(hPa) particles can be ejected from a dust sample by mere illumination with visible radiation \citep{Wurm2006b, Kocifaj2011, Kelling2011, deBeule2014}. This mechanism works independently of any wind if the illumination is strong enough. \citet{deBeule2014} found in microgravity experiments that this is caused by significant gas flow within the soil. This fact was not recognized before as thermal convection obscured this flow under gravity and only the microgravity experiments revealed the sub-surface gas flow. This flow within the dust bed is driven by thermal creep which pumps gas from a cold to a warm region if capillaries are present in a size comparable to the mean free path of the gas molecules. Mean free paths are similar to pore sizes in a dust bed at hPa pressure, hence we expect a large gas flow on Mars but an insignificant contribution under Earth's high atmospheric pressure.

A simple consideration of thermal creep under steady state conditions was proposed by \cite{Knudsen1909}.
By looking at the number of particles crossing the cross section of a capillary between two chambers with different temperatures $T_1,T_2$ a relationship between temperature and pressure can be deduced.
The number of particles crossing the section $\dot{n}$ is given by
\begin{equation}
\dot{n} \propto \frac{p v}{T}
\end{equation}
where $v$ is the thermal velocity which is proportional to $\sqrt{T}$. In equilibrium the gas flow from both sides is the same and
\begin{equation}
\frac{p_1}{p_2}=\frac{\sqrt{T_1}}{\sqrt{T_2}}.
\end{equation}
If the capillary is small, the overpressure exists as the gas only creeps along the wall and the pressure-driven back-flow cannot pass the capillary easily, but as the capillary becomes larger the flow due to the pressure difference cancels this effect.

\begin{figure*}
\centering
\includegraphics[width=0.45\textwidth]{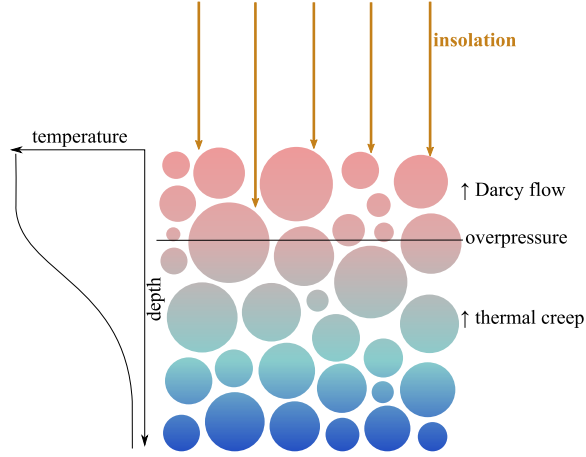}
\caption{Schematics of the temperature in an illuminated dust bed. The top layer of the dust is heated by the illumination. Gas from the bottom (or non illuminated regions) moves due to thermal creep. This leads to an overpressure just under the surface as in the topmost layer the gas can only escape by Darcy flow.}
\label{fig:temp}
\end{figure*}

Based on flow in small tubes the mass flow rate $\dot{M}$ in a dust bed can be approximated as \citep{sone1990,muntz2002}
\begin{equation}\label{eq:massflow}
\dot{M} = p_{avg} \frac{P A}{\sqrt{2 \frac{k_B}{m} T_{avg}}} \times \left(\frac{L_r}{L_x} \frac{\Delta T}{T_{avg}} Q_t - \frac{L_r}{L_x}\frac{\Delta p}{p_{avg}} Q_p \right)
\end{equation}
with $p_{avg}$ and $T_{avg}$ as the average pressure and temperature, $P$ as  porosity, $A$ as surface area, $k_B$ as Boltzmann constant, $m$ as the molecular mass of the gas, $L_r$ and $L_x$ as the radius and length of the capillaries describing the dust bed. $\Delta T$ and $\Delta p$ are the temperature and pressure differences over the capillaries, respectively. The coefficients $Q_P$ and $Q_T$ depend on the Knudsen number and describe the pressure driven (back) flow and the flow by thermal creep. The Knudsen number ($\mathrm{Kn}=\lambda/d$) is the ratio between the mean free path $\lambda$ of the gas molecules and a characteristic dimension $d$, which is $L_r$ in this case. The coefficients, $Q$, were adopted from \citet{muntz2002}. We use an analytic relation fitted to these data (K\"oster, private communication)
\begin{equation} 
Q\left(\mathrm{Kn}\right)=\frac{Q_T}{Q_P}\left(\mathrm{Kn}\right)=\frac{a_1}{(1 + \frac{a_2}{\mathrm{Kn}^{{a_{3}}}})(1 + \frac{a_4}{\mathrm{Kn}^{{a_{5}}}})},
\end{equation}
with $a_1=0.5007 \pm 0.0001$, $a_2=0.990\pm0.006$, $a_3=0.743\pm0.002$, $a_4=0.222\pm0.004$, $a_5=1.241\pm0.004$.
For "large" pores the pressure back flow dominates and no large scale net gas flow 
is present, which - again - translates to the necessity of low pressure for the effect to be significant.
A typical temperature profile of an illuminated dust bed is sketched in  figure
\ref{fig:temp}. Absolute values depend on the insolation light flux but calculated temperature profiles always show a similar structure \citep{Kocifaj2011}.

On one side the gradient below the surface implies that thermal creep pumping drives a gas flow upwards.
On the other side the uniform temperature over the top layer implies that gas flow can only be maintained if a pressure difference is present as given by Darcy's law (also included in eq. \ref{eq:massflow} as pressure term). 
\citet{deBeule2015} call this pressure supported top layer an {\it insolation activated layer}. While we do not use their results they found the activated layer thickness to be between 100 and 200\,$\rm \mu m$ in their experiments though this will vary with the environment or experimental settings and parameters. In any case their experiments confirm the existence of a sub-surface overpressure.
Balancing the pressure ($\Delta T=0$) and thermal creep ($\Delta P=0$) driven mass flows leads to an overpressure of 
\begin{equation}
\Delta p=\frac{L_{x2}}{L_{x1}+L_{x2}} p \frac{\Delta T}{T}\frac{Q_T}{Q_P},
\end{equation}
with the thickness of the layer with the temperature gradient $L_{x1}$ and of the uniform temperature layer $L_{x2}$.

We parameterize the associated lifting force $F_{al}$ on grains within the activated layer 
as
\begin{equation}
F_{al} = C_{al} I \frac{p}{T}\frac{Q_T}{Q_P}
\label{eq:activate}
\end{equation}
where $C_{al}$ is a constant which characterizes the thermal and optical properties of the dust bed. The other parameters are kept explicitly for later scaling to martian conditions. $I$ is the light flux generating $\Delta T$.
As a first order approximation we consider both to be proportional to each other.

\subsection{Aeolian Lift Model}

Aeolian entrainment is complex and cannot be calculated from first principles. Therefore the existing models of entrainment consist of semi-empirical laws formulated for Earth. To apply these on Mars they are rescaled to martian conditions (eg. \citet{greeley1980, merrison2007}).
The force which a free particle with radius $r$ experiences in a gas flow due to dynamic pressure is 
\begin{equation}
F_D=\frac{\pi}{2}C_D \rho u^2 r^2,
\end{equation}
where $u$ is the flow velocity, $\rho$ is the gas density, and $C_D$ is the drag coefficient, which depends on the flow characteristics. The force on a free particle would act along the flow direction.

At the soil / atmosphere interface the flow and its interaction with the surface leads to a gas drag lifting force opposing gravity. This force is supposed to have the same dependencies as the drag on a free particle, namely
\begin{equation}
F_L=C_L \rho u^2 r^2.
\end{equation}
It is characterized by a lifting coefficient $C_L$ which holds all unknowns and has to be a fraction of the drag coefficient $C_D$.

If the flow is not uniform across the particle diameter, the flow velocity gradient will also generate a torque, which can help to reduce the adhesion and can initiate the rolling motion of larger grains. The force related to this torque can be described by
\begin{equation}
F_T=C_T \rho u^2 r^3,
\end{equation} 
noting that $C_T$ is not dimensionless but has the unit m$^{-1}$ now.
$C_D$ is well known for spheres, the lift and torque coefficients, however, are not well constrained.

\subsubsection{Surface Flow}

The gas flow at the surface results in a the shear stress $\tau$. Related to this a shear velocity is defined as
\begin{equation}
u_*=\sqrt{\frac{\tau}{\rho}}.
\end{equation}
The flow in the vicinity of the surface follows a logarithmic profile
\begin{equation}
\label{eq:ustern}
u\left( z \right)=\frac{u_*}{\kappa} \ln \frac{z}{z_0}
\end{equation}
with the Karman constant $\kappa=0.41$ and the surface roughness $z_0$.
Gas drag depends on the friction velocity.

\subsubsection{Slip Flow Regime}

At large Knudsen numbers molecules passing a particle closely keep a ballistic trajectory (slip flow). At lower Knudsen numbers they are slowed by other molecules deflected from the particle surface but the velocity component parallel to the surface does not vanish completely. For this regime the drag can be quantified by the Stokes law with an additional slip flow correction factor depending on the Knudsen number
\begin{equation}
F_D=-6 \pi \mu r u /\mathrm{C}(\mathrm{Kn}).
\end{equation}
with $\mu$ as viscosity. The Cunningham correction factor $\mathrm{C}$ is a semi-empirical function of the Knudsen number
\begin{equation}
\mathrm{C}(\mathrm{Kn})=1+\left(q_1+q_2 \exp\left(\frac{-q_3}{\mathrm{Kn}}\right)\right)\mathrm{Kn}
\end{equation}
with $q_1=1.207$,$q_2=0.44$ and $q_3=0.78$ \citep{Rader1990}.

\subsection{Combining Aeolian Lift with the Activated Layer}

\label{sect:liftmodel}

We are interested in the dependency of the light flux needed to initiate dust motion at a given shear velocity. 
The erosion threshold at a given light flux is defined by the
balance between lifting forces $F_{L}$ and $F_{al}$ on the one side and the opposing forces gravity $F_G$ and cohesion $F_C$ or
\begin{equation}
C_L \frac{\pi}{2} \rho r^2 u_*^2 + C_{al} I \frac{p}{T}\frac{Q_T}{Q_P}= F_G+F_C.
\end{equation}
For a given grain diameter (not varied here) the lifting related to torque can effectively be included in $C_L$. We therefore did not explicitly trace this term further on but this would need to be included if particle variations would be considered. 
As critical light flux $I_{crit}$ depending on the velocity $u_*$ we therefore get
\begin{equation}
I_{crit}=\frac{F_G+F_C - C_L \frac{\pi}{2} \rho r^2 u_*^2}{C_{al} \frac{p}{T}Q}
\label{eq:fit}
\end{equation}
It has to be noted, that $Q$ depends on the Knudsen number, and $C_L$ is a function of the flow parameters. Two choices are possible: the usual one is taking $C_L$ to be a power law function of the Reynolds number. On the other hand one can use the slip flow corrected equation. The drag coefficient $C_D$ is already inserted into the slip flow equation, only the proportionality constant between drag and lift is still unknown. 

\section{Wind Tunnel Setup}
\label{sect:setup}

\subsection{Wind Tunnel Capabilities}

\begin{figure*}
\noindent
\centering
\includegraphics[width=0.9\textwidth]{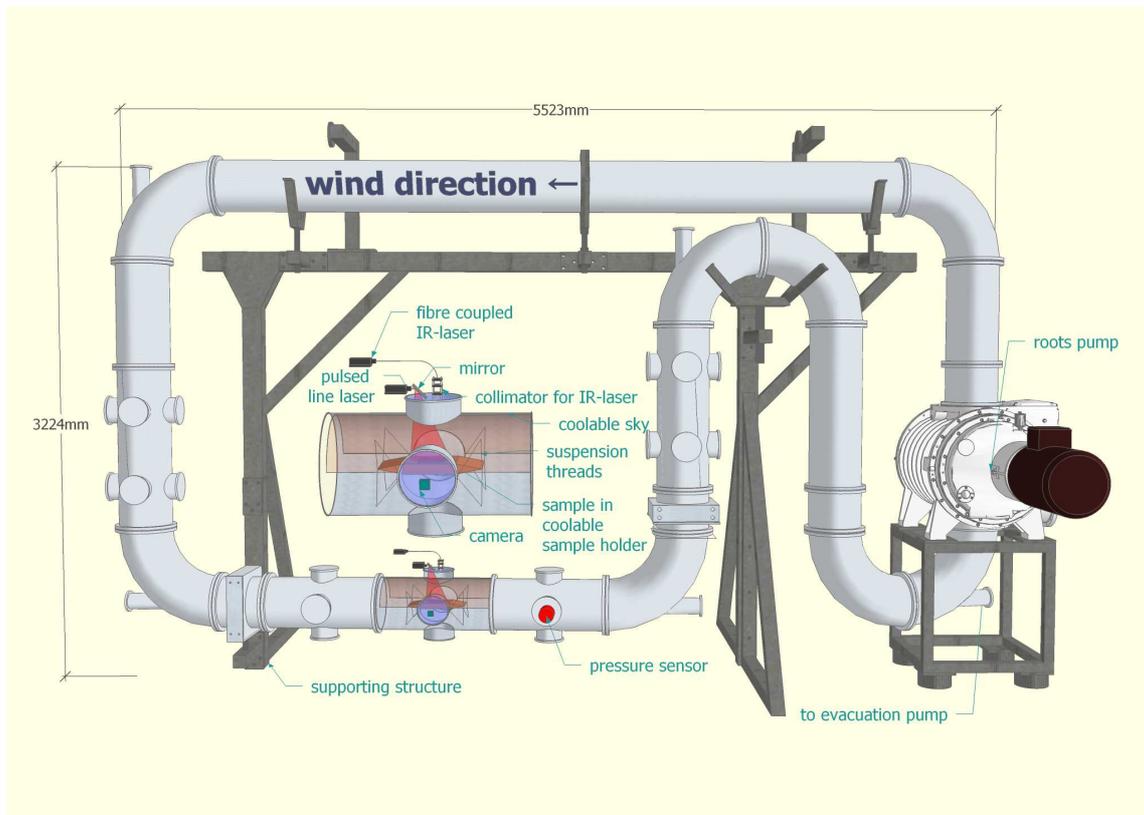}
\caption{Schematics of the wind tunnel. The wind tunnel can be evacuated by a vacuum pump. The roots pump generates the air flow. As an inset the experiment section is shown. In this section the sample is placed in a suspended sample holder and is irradiated by an infrared laser. The IR-radiation is blocked from the camera (recording with 100\,fps) by a red line-filter. Particles are visualized by additional illumination with a pulsed red laser curtain.}
\label{fig:section}
\end{figure*}

To study the influence of insolation activated lift on the threshold wind speed we
constructed a new wind tunnel based on an earlier set up by \citet{paraskov2006}. The wind tunnel has a total length of 18 m divided in several horizontal and vertical parts, including one horizontal and two vertical experimenting sections (upward, downward and horizontal wind), sketched in figure \ref{fig:section}. It consists of stainless steel tubing with an inner diameter of 32\,cm. The channel can be evacuated to and work at pressures between 10\,Pa and 1000\,Pa. The wind is provided by a roots pump circulating up to 8\,$\rm m^3/s$. This translates to a wind speed averaged over the cross section up to 100\,m/s.

\subsection{Experimental Setup}

The experiments were carried out in the horizontal 2.4\,m long experimenting section.

A sample holder with a simple wing shape is suspended in the middle of the channel by several threads. A dust sample with a depth of 2.7\,cm and a square surface of 92\,cm$^2$ can be prepared and inserted. 
The dust sample is placed in the center of the channel as sketched in figure \ref{fig:section}. 

A cw infrared laser (1064\,nm) at varying light flux illuminated the dust bed. To visualize the (airborne) particles an additional red laser curtain was implemented. Dust particles are detected by a long term recording camera running at a rate of 100 frames per second, with maximal exposure time (9994\,\textmu s). The heating of the dust bed is dominated by the IR laser while the heating due to the laser curtain is insignificant. As its scattered light would overexpose the images a line filter in front of the camera removes the IR and only the red laser curtain is visible. The laser curtain was pulsed with 3 short pulses of 164\,\textmu s duration with 176\,\textmu s delay in between followed by a 2.45\,ms break, so every particle shows up in the camera pictures as a distinct pattern from which the speed and acceleration could be calculated.

In this work we only distinguish whether lift occurs in a given experiment. A measurement of lifting rates is outside the scopes of this paper. The infrared laser allows for an intensity much higher than solar, which is not meant to simulate insolation on Mars but improves the determination of the dependence of the threshold light flux on friction velocity at martian values.

\subsection{Sample Preparation}

\begin{figure}
\noindent
\centering
\includegraphics[width=0.45\textwidth]{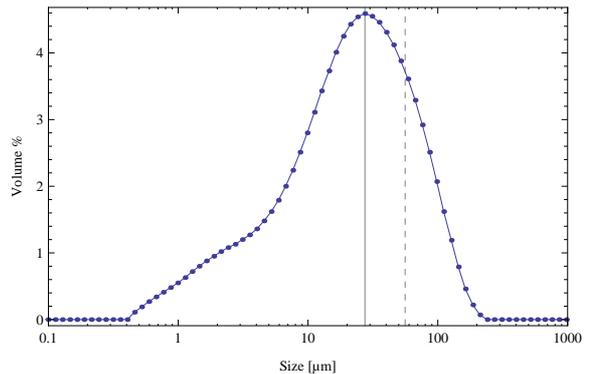}
\caption{Volume size distribution of the dust samples. Peak of the distribution is at 27\,\textmu m and 50\,\% of the dust is smaller than 56\,\textmu m.}
\label{fig:sizedistribution}
\end{figure}

As dust sample material we used Mojave Mars Simulant \citep{peters2008}. It consists of crushed basaltic rock. It is more angular than the aeolian weathered dust expected on Mars. As it is not weathered by the influence of aqueous alteration it has a low volatile content of only up to 4\,\%. It was additionally treated by heating it to 400\,\textdegree C for 6\,hours, to reduce the volatile content further.
The size distribution of particles in the dust bed was measured with a commercial particle sizer (Malvern Mastersizer 3000) and is shown in figure \ref{fig:sizedistribution}. \citet{peters2008} consider this dust to be representative for the mechanical properties of the fine grained material on Mars, as the inventory of sub micron particles is still included in the sample.
It has also spectral characteristics and material composition which are close to those of Mars \citep{peters2008}. Therefore, it was deemed as a suitable sample material for a first study here.

The dust was sieved into the sample holder to achieve a loose settling of the grains and prevent compaction which may alter the physical properties of the surface. In this process some very loosely bound aggregates or grains may reside at the surface, and those particle are lifted easily.
We did not intend to study the threshold speeds for erosion of some particles which are exceptionally susceptible to pick up but were interested in the more constant entrainment thresholds. As procedure to remove these particles and reproduce similar conditions for all experimental runs the sample was passivated before the measurement. This was achieved by a >30\,s gas flow with a higher speed than used in the measurement. This removed all particles that are only loosely bound to the surface and allows the determination of a threshold for continuous entrainment afterwards. This is not necessarily the most natural soil expected on Mars where atmospheric dust on the order of 1\,$\rm \mu m$ might settle on top of a coarser bed. However, it is a well defined way of preparing the dust bed and warrant similar conditions for each experiment run. 
The ambient pressure was set to 320\,Pa, 630\,Pa and 930\,Pa, with uncertainties of 5\,\% during the experiments, and air was used as working gas.

\begin{figure*}
\noindent
\centering
\includegraphics[width=0.8\textwidth]{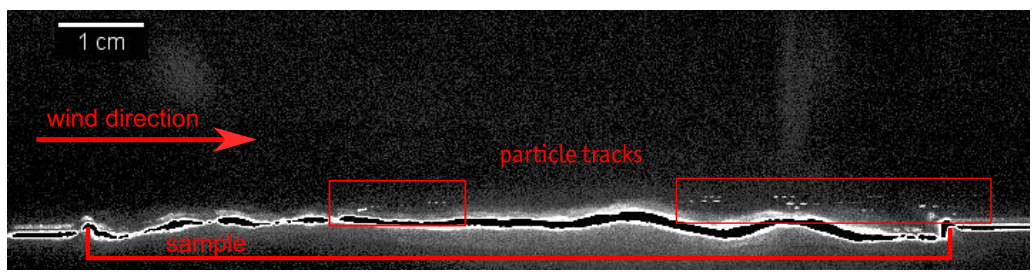}
\caption{Example of a processed sequence of images showing lifted particles.}
\label{fig:frame}
\end{figure*}

In each experimental run a given sample is observed at different flow velocities and light fluxes. First the lowest light flux is set and the wind speed is increased. Then speed is decreased again, the light flux is increased and so on. This gives a matrix which samples the 2d space of wind speed and light flux.    
Each measurement consists of 2750 frames (27.5 seconds of data acquisition at 100\,fps). These frames are processed and overlayed to produce one image for each measured sequence which shows all entrained particle~-- if present. This way a decision can be taken for each image sequence if particles are picked up within the laser curtain or not. An example is given in figure \ref{fig:frame}.

\section{Results}
\label{sect:results}

The data are shown in table \ref{tab:measurements}. Here, the flow is characterized in terms of friction velocity as described in the next section.

\begin{table*}
\caption{Experiment results for three used pressures. The number of experiments (out of 6), where particle lift occurred, is shown depending on light flux (first column) and friction velocity. The thresholds were determined where at least two measurements showed erosion (or the previous measurement already showed erosion). The thresholds are shown in bold. Towards lower pressure the rates become smaller and the threshold is less visible, therefore the data at 205 Pa was not used for fitting but is shown for completeness.} 
\centering
\begin{tabular}[c]{|l|c|c|c|c|c|}
\hline
$u_*$ in m/s & 1.31&1.53&1.68&1.94&2.32 \\ \hline
I in kW/m$^2$ &  \multicolumn{5}{c|}{2.05\,mbar} \\ \hline
0.39 &0&0&0&0&1\\ \hline
1.83 &0&0&0&1&0\\ \hline
3.80 &0&0&0&{\bfseries 2}&{\bfseries 1}\\ \hline
5.50 &1&{\bfseries 3}&0&2&1\\ \hline
7.28 &{\bfseries 3}& 2&1&4&3\\ \hline
\hline
$u_*$ in ${\mathrm{m}}/{\mathrm{s}}$ & 1.1&1.31&1.53&1.68&1.94 \\ \hline
I in ${\mathrm{kW}}/{\mathrm{m}^2}$ &  \multicolumn{5}{c|}{3.2\,mbar}  \\ \hline
0.62 &						0 &0 &1 &1 &{\bfseries 2}             		\\  \hline
2.43 &						0 &0 &0 &{\bfseries 3} &6            			\\  \hline
5.03 &						0 &{\bfseries 3} &{\bfseries 3} &4 &6 		\\ \hline
7.25 &			      0 &3 &4 &4 &6                        		\\ \hline
8.91 &						1 &1 &2 &5 &6 													\\ \hline \hline
$u_*$ in ${\mathrm{m}}/{\mathrm{s}}$ & 0.88& 1.1 &1.31& 1.53 &1.68 \\ \hline
I in ${\mathrm{kW}}/{\mathrm{m}^2}$ &  \multicolumn{5}{c|}{6.3\,mbar} \\ \hline
0.62 &0 &0 &1 &1 &{\bfseries 2} 			\\  \hline
2.43 & 0 &0 &0 &{\bfseries 3} &6 			\\  \hline
5.03 &0 &{\bfseries 3} &{\bfseries 3} &4 &6 			\\ \hline
7.25 &0 &3 &4 &4 &6				\\ \hline
8.91 &1 &1 &2 &5 &6				\\ \hline \hline
$u_*$ in ${\mathrm{m}}/{\mathrm{s}}$ &  0.88& 1.1 &1.31& 1.53 &1.68 \\ \hline
I in ${\mathrm{kW}}/{\mathrm{m}^2}$ & \multicolumn{5}{c|}{9.3\,mbar} \\ \hline
0.62 &	0&0&0&0&{\bfseries 6}			\\  \hline
2.43 &	0&0&0&{\bfseries 4}&6			\\  \hline
5.03 &	0&1&{\bfseries 3}&5&6			\\ \hline
7.25 &	0&{\bfseries 3}&5&6&6			\\ \hline
8.91 &	0&4&5&6&6			\\ \hline
\end{tabular}\label{tab:measurements}
\end{table*}

\subsection{Boundary Layer Profile}

\begin{figure}
\centering
\includegraphics[width=0.48\textwidth]{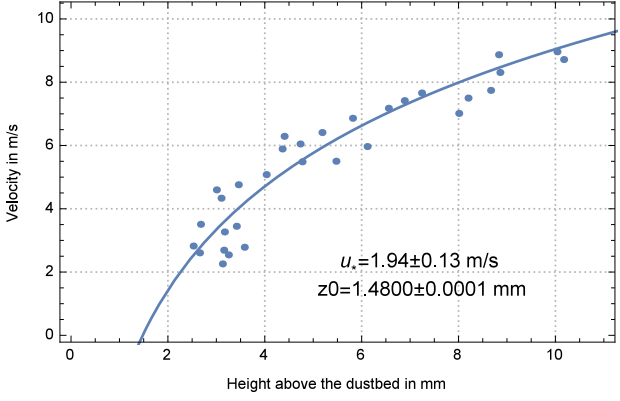}
\caption{Velocity over height for grains detached at largest wind speed used at 320 Pa. A fit to the data according to equation \ref{eq:ustern} is shown.}
\label{fig:velofit}
\end{figure}

To determine the boundary layer profile, the velocity of the entrained grains was measured in dependence of the height. We used one data set at the highest wind velocity used at 320\,Pa. Here a large number of particles could be detected to allow a flow characterization. The last visible occurrence of each particle was used as a data point. There the particles were already well coupled to the gas flow and hence trace its velocity.  The boundary layer flow is shown in figure \ref{fig:velofit}.
The data were fitted to equation \ref{eq:ustern} to obtain surface roughness and friction velocity. The friction velocities at other settings was calculated assuming that $u_{*}\propto u_{mean}$. The average flow velocity in the channel $u_{mean}$ is known for the pump system.

\subsection{Detachment}

All measurements are summarized in table \ref{tab:measurements}. This shows the number of runs where lifted particles were detected depending on wind speed and light flux, 6 runs were made for every combination thereof. It is clearly visible that the threshold speed is reduced with increasing illumination and decreases with decreasing pressure. At 205\,Pa the determination of the threshold is ambiguous and this data was not used for any fitting but is shown for completeness.  

The data have to match the dependency of equation \ref{eq:fit}. As $Q_T/Q_P$ depends on the Knudsen number, the mean free path $\lambda$ was calculated but the particle size $d$ was left as a fit parameter. For the lifting coefficient $C_L(\mathrm{Re})$ two approaches were used and compared:
The ansatz in the literature is using mostly Newtonian drag and assuming a power law function for $C_L$. A second ansatz was also tested. For the given particle size and velocities the appropriate drag should be described by the slip flow regime, therefore this dependency was also used.

\begin{figure}
\centering
\includegraphics[width=0.48\textwidth]{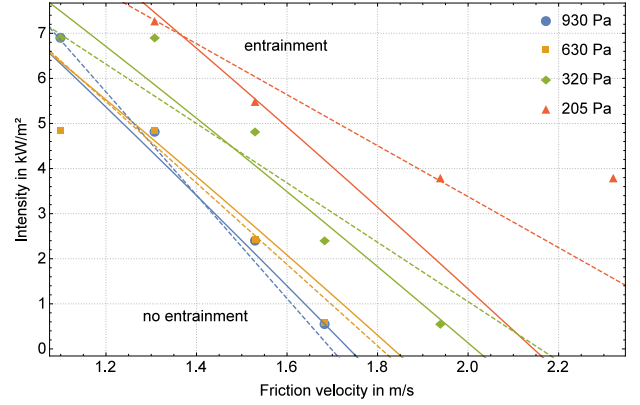}
\caption{Model fits to the entrainment thresholds obtained at 930\,Pa, 630\,Pa and 320\,Pa. The data at 205\,Pa was not used for fitting. The Newtonian drag fit is shown as solid lines and the slip flow fit as dashed lines. Both models have a similar fit quality. For details we refer to the text.}
\label{fig:threshold}
\end{figure}

For the Newtonian drag a power law $C_L(\mathrm{Re}) =\mathrm{const} \cdot \mathrm{Re}^\gamma$ was assumed, where the constant is combined with the other parameters in front of $u_*^2$ as new parameter $\tilde{\alpha}_N$. Two constants $u_{*,0}=1$\,m/s and $p_0$=600\,Pa were inserted to avoid confusion with the units. The parameters $(F_G+F_C)/C_{al}T$ were replaced by one fitting parameter $F_{GC,N}$. In total this gives the function
\begin{equation}
\scalebox{0.85}{$
I_{crit}\left(u_*,p\right)=\frac{p_0}{p}Q(\frac{\lambda}{d_N})\left(-\tilde{\alpha}_N \frac{p}{p_0}\left(\frac{u_* p}{p_0 u_{*,0}}\right)^\gamma u_*^2 + F_{GC,N}\right),
$}
\label{eq:newt}
\end{equation} 
which was fitted to the data obtained for all pressures.
The fit to our experimental data can be seen in figure \ref{fig:threshold} and the four parameters obtained are: $\tilde{\alpha}_N=0.59 \pm 0.24 $\,kWs$^2$/m$^4$, $\gamma=-0.83\pm0.07$, $d_N=32\pm11$\,\textmu m, and $F_{GC,N}=1.20 \pm 0.46$\,kW/m$^2$.
The parameter $d_N=$32\,\textmu m is consistent with the particle size. $\gamma$ gives the dependence on the Reynolds number which is $\propto p^x$ with -1<x<0. Therefore, the value for $\gamma$ is plausible. 

For the slip flow the fit function is given by
\begin{equation}
\scalebox{0.85}{$
I_{crit}\left(u_*,p\right)=\frac{p_0}{p}Q(\frac{\lambda}{d_C})\left(-\tilde{\alpha}_C d \frac{u_*}{\mathrm{C}(\lambda/d_C)}+F_{GC,C}\right),
$}
\label{eq:cun}
\end{equation} 

with the three fit parameters being $\tilde{\alpha}_C=0.0086 \pm 0.0026 $\,kWs/m$^4$,$d_C=56\pm7$\,\textmu m, and $F_{GC,C}=0.70 \pm 0.1$\,kW/m$^2$.

\subsection{Scaling to Mars}

The laboratory results have to be scaled to martian conditions to account for differences in insolation, gravity, temperature and gas composition. 
The differences between laboratory conditions and Mars are summarized in table \ref{tab:rescaling}. 

\begin{table*}
\caption{Comparison of relevant laboratory and martian parameters, the ratio is the value on Earth divided by the martian value.}
\centering
\begin{tabular}{|lc r|c|c|r|}
\hline
Quantity &Unit& Symbol & Earth & Mars & ratio [\%] \\ \hline
gravity &[m/s$^2$] &g& 9.81 & 3.69 & 37.6 \\
pressure & [hPa] &p& 1013 & 6 & 0.6 \\
temperature & [K] &T& 295 & 218 & 73.9 \\ \hline
major Atmospheric component &  & & N$_2$ & CO$_2$ & - \\
 molecular mass & [u] &m& 28.96 & 44.01 & 152 \\
molecular cross section & [10$^{-20}$\,m$^2$] &$\sigma$ & 11.34 & 9,08 & 80.1 \\
max insolation& [kW/m$^2$] &$I_{max}$& 1.412 & 0.717 & 50 \\
\hline
\end{tabular}
\label{tab:rescaling}
\end{table*}

Every parameter of equations \ref{eq:newt} and \ref{eq:cun} can be recaled by multiplying it by a scaling factor $f$. For the Newtonian drag we use the following: $Q_T/Q_P$ is a function of the Knudsen number. The Knudsen number scales with the mean free path $\lambda=(\sigma n)^{-1}$ where sigma is the molecular cross section and $n=p/k_b T$ the number density. Therefore is can be calculated that $\lambda \propto T/\sigma$ and $f_\lambda=0.923$.  The insolation induced lifting force scales with $1/T$, and $f_{lift}=1.35$. The drag force scales with the density $\rho \propto p \cdot m/T$ where $m$ is the molecule mass and the scaling factor takes the value $f_\rho=2.06$. The drag coefficient scales with the Reynolds number, which again scales with the density. These factors can be inserted to rescale the fit, giving 
\begin{equation}
\scalebox{0.85}{$
I_{crit}\left(u_*,p\right)=\frac{p_0}{p}Q(\frac{\lambda f_\lambda}{d_N})\left(-\tilde{\alpha}_N \frac{f_\rho}{f_{lift}} \frac{p}{p_0}\left(\frac{u_* f_\rho p}{p_0 u_{*,0}}\right)^\gamma u_*^2 + \frac{f_{F_{GC}}}{f_{lift}}F_{GC,N}\right).
$}
\end{equation} 
Only $f_{F_{GC}}$ is difficult to obtain as the degeneracy between gravity and cohesion can not be disentangled from this dataset. Therefore the two limiting cases will be treated: If $F_{GC}$ is dominated  by gravity $f_{F_{GC}}=1/3$ as gravity on Mars is only 1/3 compared to Earth. If cohesion dominates $f_{F_{GC}}=1$ as cohesion stays the same. The real value will be somewhere in between.
The Cunningham fit function and the other parameters can be scaled accordingly if slip flow is considered instead of Newtonian flow.

\begin{figure*}
\centering
\includegraphics[width=0.6\textwidth]{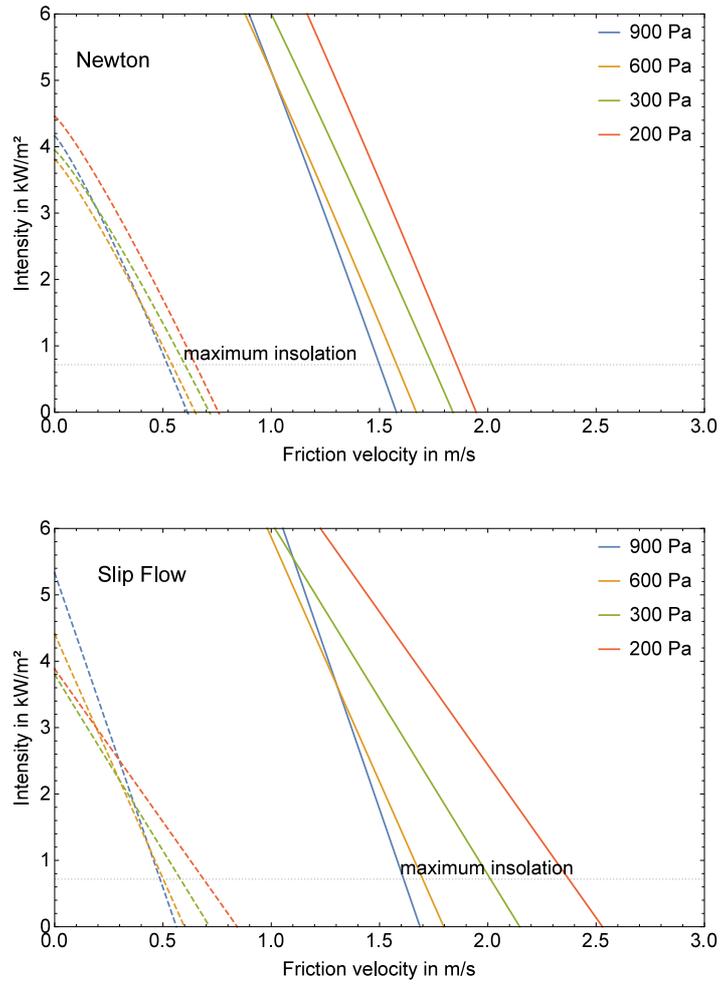}
\caption{Threshold conditions rescaled to martian environment; solid lines: cohesion
dominates over gravity, dashed lines: gravity dominates over cohesion; note the different scales on the x-axis.}
\label{fig:scaled}
\end{figure*}

For the maximum insolation on Mars of 717\,$\rm W/m^2$ we get threshold friction velocities compiled in table \ref{tab:rescaled}. Compared to the calculated threshold friction velocity with no insolation the reduction lies between 4\,\% and 19\,\%, depending on wheter the holding force is cohesion or gravity. The rescaled fit is shown in figure \ref{fig:scaled}.

\begin{table*}
\caption{Threshold friction velocities under different conditions on Mars}
\centering
\begin{tabular}{|c||c|c|c||c|c|c||c|c|c|}
\hline
\multicolumn{10}{|c|}{Newton} \\ \hline
 & \multicolumn{3}{c||}{lab values} & \multicolumn{3}{c||}{$f_{F_{GC}}=1$}& \multicolumn{3}{c|}{$f_{F_{GC}}=1/3$} \\ \hline
 p & $u_{*,I=0}$ & $u_{*,I_{max}}$ & red. & $u_{*,I=0}$ & $u_{*,I_{max}}$ & red.& $u_{*,I=0}$ & $u_{*,I_{max}}$ & red. \\ 
$$[Pa] & [m/s] & [m/s] & [\%] & [m/s] & [m/s] & [\%] & [m/s] & [m/s] & [\%] \\ \hline
 205 & 2.14 & 2.07 & 3.6 & 1.94 & 1.85 & 4.6 & 0.75 & 0.65 & 14.1\\
 320 & 2.02 & 1.93 & 4.2 & 1.81 & 1.73 & 5.2 & 0.78 & 0.59 & 16.1\\
 630 & 1.83 & 1.75 & 4.4 & 1.66 & 1.57 & 5.3 & 0.64 & 0.54 & 16.3\\
 930 & 1.74 & 1.67 & 4.0 & 1.57 & 1.50 & 4.8 & 0.61 & 0.52 & 14.9\\
  \hline  \hline  \multicolumn{10}{|c|}{Slip Flow} \\ \hline
 & \multicolumn{3}{c||}{lab values} & \multicolumn{3}{c||}{$f_{F_{GC}}=1$}& \multicolumn{3}{c|}{$f_{F_{GC}}=1/3$} \\ \hline
 p & $u_{*,I=0}$ & $u_{*,I_{max}}$ & red. & $u_{*,I=0}$ & $u_{*,I_{max}}$ & red.& $u_{*,I=0}$ & $u_{*,I_{max}}$ & red. \\ 
$$[Pa] & [m/s] & [m/s] & [\%] & [m/s] & [m/s] & [\%] & [m/s] & [m/s] & [\%] \\ \hline
205 & 2.60 & 2.47 & 4.9 & 2.50 & 2.35 & 6.0 & 0.83 & 0.68 & 18.5\\
320 & 2.16 & 2.05 & 5.0 & 2.10 & 1.97 & 6.1 & 0.70 & 0.57 & 18.8\\
630 & 1.81 & 1.73 & 4.4 & 1.78 & 1.67 & 5.2 & 0.59 & 0.50 & 15.9\\
930 & 1.70 & 1.63 & 3.7 & 0.56 & 0.49 & 4.3 & 0.56 & 0.49 & 13.2\\
 \hline
 \end{tabular}
 \centering
 \label{tab:rescaled}
 \end{table*}

Under our laboratory conditions Newtonian flow and corrected slip flow models differ only significantly at very low pressure. Due to the limited capability of particle lifting though, the measurements do not allow to distinguish between both. Scaled to Mars the differences get somewhat larger. However, it should be noted that the relative reduction is rather robust independent of the drag law considered.

\section{Discussion}

The manually prepared dust beds and artificial illumination can provide a model of a planetary surface, but not all parameters can be mimicked.
The Reynolds number at 600\,Pa and\,30 m/s mean velocity for our wind channel is 3200. The 30\,m/s correspond to a friction velocity of 1.86\,m/s. This means that the flow is slightly above the critical Reynolds number for some measurements~-- namly the highest velocity at 320\,Pa and the two highest at 205\,Pa~-- and hence should be prone to become turbulent. At lower speeds for higher light flux it drops below the critical Reynolds number though. We do not expect that the transition influences the systematics of the measurements, as the measurement at 205\,Pa was not taken into account.

The IR light source illuminated an area of  1.77\,cm$^2$ and was on during the experiment run lasting about half an hour. This implies a spatially confined and time dependent temperature variation over the course of each experiment. Gas flow and pressure build up below the surface might therefore vary over time. To what extend this will affect the data is unclear. 
However, as the expected dependence fits the data rather well we do not expect the systematics to change strongly.

Care has been taken to increase the light flux during a measurement, as decreasing the light flux can otherwise result in particle ejection. This is a related lifting mechanism e.g. studied in \citet{Kelling2011}. It is based on a different temperature profile within the soil developing after a decrease of the illuminating flux. This might be of high importance for particle lift e.g. in dust devils providing shadowing on short time scales or a shadow cast by a surface feature moving over the ground. However, this is not the topic of this work, but shall be examined in the future. 
Here we are interested in the supporting lift for a continuous illumination in face of a wind flow. By increasing the light flux during the experiments we avoid the occurrence of such particle lift.
If a dust storm sets in, its opacity will hinder the insolation, for a sharp cut in optical density this might add to the lifting. But as saltation was not present at this experiment~-- as the saltation length is larger than the experiment section~--, the evolution of a dust storm can not be predicted here.

There is some variability between the samples, where some samples are more readily eroded than others but this has to be expected for dust samples prepared manually.
This raises the question how a natural dust surface would behave if small dust particle in the micrometer range would settle on top of a dust bed. As this influences the gas flow and pressure profile in the dust bed it might have some
influence on particle lifting. The permeability of the top layer gets smaller which requires a higher pressure to force the gas flow through. This would ease particle lift. This has to be studied in future experiments.

Note that we observe no signs of electrical charging effects here and do not consider it to play a role in the context of this study, as the amount of charge will not be influenced by the infrared radiation, and cannot provide a relative reduction in respect to the illumination.

Direct predictions for the dust entrainment activity is not straight forward, as a multitude of properties have to be taken into account. The highest intensity available is at perihelion at sub-solar latitude. But temperature and gas density play major roles. Therefore, dust activities should be tied to certain regions on Mars, which is consistent with observations. This correlation was already found by \cite{Wurm2008}, who investigated the illumination effect alone.

As the top layer insolation activation depends on numerous parameters the absolute wind speed reduction is variable. Pore sizes and particle size distribution are only two examples to be varied.  We do not deem it possible to quantify the effects of all variations here but several options would increase the activation effect.
Further experiments would have to be carried out to quantify the influence of the different parameters.

However, these first experiments imply that insolation activation can put significant tension on a top soil layer under the low pressure conditions of Mars and enhance particle lifting by wind or within dust devils.
In this first systematic analysis of insolation aided wind erosion we found a reduction of the threshold speed of about 4\,\% to 19\,\% for martian conditions. As it is unlikely that we accidentally probed the most susceptible conditions, this number is expected to be higher for some parameter combinations. Therefore, the insolation activated layer might strongly enhance particle lifting at lower wind speeds.

\section{Conclusion}

In this work we present laboratory measurements on the reduction of the threshold friction velocity necessary for lifting dust if the dust bed is illuminated. Insolation driven thermal creep can activate a dust layer by adding a sub-surface overpressure. This supports lift imposed by wind. Our first measurements imply that the insolation of the martian surface (with 717 W/m$^2$) can reduce the entrainment threshold velocity between 4\,\% and 19\,\% for the conditions sampled by the experiments. An insolation activated soil might therefore provide additional support for aeolian particle transport at low wind speeds.

The reduction of up to 19\,\% is one step to bridge the gap between observed thresholds and theoretical predictions though it might not explain all discrepancies. It still has to be investigated how this mechanism works out with sand sized particles or a broad size distribution of particles. Chaning pore size and adding particles susceptible to lift might be promising in reducing the gap further.

\subsection*{Acknowledgments}
Data (processed images) is available on request from the corresponding author. 
This work is funded by the Deutsche Forschungsgemeinschaft. We thank P. Whelley and three anonymous reviewers for a thorough review.
This work is originally published in JGR:Planets \doi{10.1002/2015JE004848}

\begingroup
\raggedright
\sloppy
\printbibliography
\endgroup

\end{document}